\documentclass[titlepage,a4paper,11pt]{article}
\usepackage[utf8]{inputenc}
\usepackage{amssymb}
\usepackage{amsmath}

\usepackage{graphicx}
\usepackage{epstopdf}
\usepackage{booktabs}
\usepackage{hyperref}
\hypersetup{
    colorlinks = true,
    citecolor=blue,
    linkcolor=black
}

\def\LL{{\cal L}}

\def\NN{{\cal N}}

\def\SS{{\cal S}}

\def\Tr{{\rm {Tr}}}

\def\beq{\begin{equation}}
\def\eeq{\end{equation}}
\newcommand{\bea}{\begin{eqnarray}}
\newcommand{\eea}{\end{eqnarray}}
\def\bal{\begin{align}}
\def\eal{\end{align}}

\begin{document}
\begin{titlepage} 
\begin{flushright}
CCQCN-2015-103\\
CCTP-2015-20
\vskip -1cm
\end{flushright}
\vskip 3cm
\begin{center}
\font\titlerm=cmr10 scaled\magstep4
    \font\titlei=cmmi10 scaled\magstep4
    \font\titleis=cmmi7 scaled\magstep4
    \centerline{\bf \LARGE \titlerm 
    A Lagrangian for self-dual strings}
\vskip 1cm
{Vasilis Niarchos}\\
\vskip 0.5cm
\medskip
{\it Crete Center for Theoretical Physics and}\\
{\it Crete Center for Quantum Complexity and Nanotechnology}\\
{\it Department of Physics, University of Crete, 71303, Greece}\\
\medskip
{niarchos@physics.uoc.gr}

\end{center}
\vskip .5cm
\centerline{\bf Abstract}

\baselineskip 20pt
%

\vskip .5cm 
\noindent
We propose a Lagrangian for the low-energy theory that resides at the $(1+1)$-dimensional 
intersection of $N$ semi-infinite M2-branes ending orthogonally on $M$ M5-branes 
in ${\mathbb R}^{1,2} \times {\mathbb C}^4/{\mathbb Z}_k$ (for arbitrary positive integers $N,M,k$). 
We formulate this theory as a 2d boundary theory with explicit $\NN=(1,1)$ supersymmetry 
that contains two superfields in the bi-fundamental representation of $U(N)\times U(M)$ 
interacting with the $(2+1)$-dimensional $U(N)_k\times U(N)_{-k}$ ABJM Chern-Simons-matter
theory in the bulk. We postulate that the boundary theory exhibits in the deep infrared supersymmetry 
enhancement to $\NN=(4,2)$, or $\NN=(4,4)$ depending on the value of $k$. Arguments in favor of the 
proposal follow from the study of the open string theory of a U-dual type IIB Hanany-Witten setup.
To formulate the bulk-boundary interactions special care is taken to incorporate all the expected 
boundary effects on gauge symmetry, supersymmetry, and other global symmetries.

\vfill
\noindent
\end{titlepage}\vfill\eject

\setcounter{equation}{0}

\pagestyle{empty}
\small
\vspace*{-0.7cm}
\tableofcontents
\normalsize
\pagestyle{plain}
\setcounter{page}{1}

\section{Introduction} 
\label{setups}

Since M2-branes can end on M5-branes it has long been suspected that the M5-brane theory 
is a still illusive six-dimensional non-critical string theory. The strings of this theory are 
charged under a self-dual three-form field strength, hence they are frequently referred to as 
self-dual strings. When the M5-branes are coincident the theory is non-abelian and intrinsically 
strongly coupled. As a result, it has proven a very hard problem to formulate this string theory 
and to extract directly information about the quantum physics of M5-branes.

Clearly, the two-dimensional intersection of M2-branes ending on M5-branes is at the heart of this problem. 
It would be useful to understand precisely the degrees of freedom that reside on this intersection and how
they interact with the three-dimensional and six-dimensional bulk on the M2 and M5-branes respectively. 
It is sensible to analyse this problem first in a symmetric configuration, e.g.\ the half-BPS
configuration of $N$ coincident M2-branes (extended along the half-plane $x^2>0)$ 
that end orthogonally on $M$ coincident M5-branes
\beq
\label{setupaa}
	\begin{array}{r c c c c c c c c c c c c}
	N ~ M2 & ~:~ & 0  & 1 & 2_+ &  &  &  &  &  &  &  &
	\\
	M ~ M5  & ~:~ & 0  & 1 &  &  &  &  &  & 7 & 8 & 9 & 10
	\end{array}
\eeq

In flat space the two-dimensional theory at the intersection enjoys large $\NN=(4,4)$ supersymmetry.
To date there has been very limited information about this theory. Let us summarize quickly some of the most 
prominent developments that are pertinent for this paper.

From the M5-brane point of view the orthogonal M2-branes can be viewed as a string soliton spike. The first successful
description of this soliton (as an M-theory BIon) was given by Howe, Lambert and West \cite{Howe:1997ue} 
for a single M5-brane. A similar analysis for coincident M2 and M5-branes in the large-$N, M$ limit was performed in 
\cite{Niarchos:2012pn} using a holographic supergravity analysis based on the blackfold approach
\cite{Emparan:2009cs,Emparan:2009at}. 
A noteworthy result of that work was a specific prediction for the leading behavior of the 
central charge $c$ of the putative two-dimensional superconformal
theory at the intersection \cite{Niarchos:2012cy,Niarchos:2013ia}
\beq
\label{setupab}
c \sim \frac{N^{\frac{3}{2}}}{\sqrt{\lambda}}+\ldots 
\sim \frac{M^3}{\lambda^2}+\ldots
\eeq
in a 't Hooft-like limit where $N,M\gg 1$ with the ratio $\lambda = \frac{M^2}{N}$ fixed and large. The dots indicate
subleading terms in a $1/\lambda$ expansion. The appearance of the powers $N^{\frac{3}{2}}$ and $M^3$ in the 
two expressions on the rhs of \eqref{setupab} is suggestive of a close relation to the well-known scaling of massless 
degrees of freedom of M2-branes $(N^{\frac{3}{2}})$, and M5-branes $(M^3)$.  
Different expressions for $c$ in other regimes were derived in 
\cite{Berman:2004ew} using anomaly considerations in the Coulomb branch of the M5-brane 
theory.\footnote{For an interesting observation on the role of self-dual string junctions
in the Coulomb phase of the ADE 6d (2,0) superconformal fields theories and the problem of the $M^3$ scaling
of the massless degrees of freedom on the M5-branes see \cite{Bolognesi:2011rq}.}

Let us note in passing that an exact fully localized half-BPS supergravity solution that describes 
the M2-M5 configuration \eqref{setupaa} in flat space is currently not known. For an older attempt to this problem
we refer the reader to \cite{Lunin:2007mj}. A more recent analysis of AdS solutions that are presumably 
near-horizon limits of M2-M5 configurations was performed in \cite{Bachas:2013vza}. It would be interesting
to distill further information about the quantum properties of the M2-M5 intersections from such 
asymptotically AdS solutions in supergravity.

There have also been several attempts to analyze the field theory of the intersection 
\eqref{setupaa} from the viewpoint of the M2-branes. 
An M-theory generalization of the Nahm equations for the BIon was proposed by Basu and Harvey in
\cite{Basu:2004ed}. Subsequently, with the advent of the ABJM model \cite{Aharony:2008ug}, the low-energy theory
on $N$ M2-branes was formulated as a $U(N)\times U(N)$ Chern-Simons-matter theory with 
explicit $\NN=6$ supersymmetry. 
The properties of semi-infinite M2-branes ending on $M$ M5-branes are captured from this perspective by
appropriate boundary conditions and/or appropriate boundary degrees of freedom in the ABJM model on 
a half-plane. 

The effects of boundaries in supersymmetric Chern-Simons-matter theories were considered 
by several authors. A formulation of the boundary problem in the context of the M2-M5 system 
in terms of supersymmetric boundary conditions was put forward in \cite{Berman:2009xd}. 
Other authors considered an alternative formulation that employs 
suitable boundary degrees of freedom. With emphasis on the boundary effects on supersymmetry 
Ref.\ \cite{Berman:2009kj} considered possible 
boundary interactions in $\NN=2$ Chern-Simons-matter theories using the technology of \cite{Belyaev:2008xk}.
A different set of boundary interactions, that emphasized the role of gauge symmetry, was considered in 
\cite{Chu:2009ms,Faizal:2011cd}. Although both approaches in this direction are technically relevant for the M2-M5 
system, their precise implementation to this problem has been obscure, because a clear M-theory guide 
to the boundary degrees of freedom and interactions that are needed to describe the M2-M5 system was mostly lacking. 
A specific proposal towards the resolution of these issues will be the main contribution of this paper.

Finally, in more recent developments it has proven useful to consider configurations of intersecting M2 and
M5-branes with compactified worldvolumes. In this context a computation of the elliptic genus of M2-branes 
suspended between parallel M5-branes was performed in \cite{Haghighat:2013gba,Haghighat:2013tka,Kim:2014dza},
and \cite{Hosomichi:2014rqa}. Ref.\ \cite{Gomis:2014eya} considered M5-branes wrapped around punctured 
Riemann surfaces. In this setup the M2-branes realize surface operators in four-dimensional $\NN = 2$ field theories.

\subsection*{Main contribution and brief summary of the paper}

The approach we take in this paper is particularly simple. The successful formulation of the 
low-energy theory on multiple M2-branes as supersymmetric Chern-Simons-matter theory, \cite{Aharony:2008ug}, 
relied on a U-dual description of M2-branes as D3-branes suspended between appropriate stacks of 
5-branes in a type IIB Hanany-Witten setup. In section \ref{lift} we describe how to incorporate an extra stack of $M$ 
D5-branes in this setup, where $N$ D3-branes can end on a half-BPS $(1+1)$-dimensional boundary. We show that the 
new configuration lifts in M-theory to the M2-M5 system of interest probing a ${\mathbb C}^4/{{\mathbb Z}_k}$
orbifold singularity. For $M=0$ D5-branes our setup reduces to the well-known brane configuration of 
\cite{Aharony:2008ug}.

In section \ref{qft} we use the type IIB setup to read off the spectrum and interactions at the D3-D5 boundary.
We find that the massless boundary degrees of freedom that arise in the D3-D5 open string theory are two sets of 
2d $\NN=(1,1)$ supermultiplets in the bi-fundamental representation of the $U(M) \times U(N)$ group.
Using a formulation with explicit $\NN=2$ supersymmetry in the three-dimensional bulk 
we present a 2d boundary theory that exhibits 
$\NN=(1,1)$ supersymmetry. Precise bulk-boundary interactions of this theory are proposed using
the recent results of Ref.\ \cite{Armoni:2015jsa}, that is building on the previous works 
\cite{Belyaev:2008xk,Chu:2009ms,Faizal:2011cd}. Analyzing the symmetries of the postulated 
action and the symmetries of the underlying brane setup we postulate that for generic Chern-Simons level $k>2$ 
the bulk-boundary theory flows in the deep infrared to a fixed point with the expected 2d $\NN=(4,2)$ supersymmetry.
We anticipate a further enhancement of the boundary supersymmetry for the special value $k=1$ to 
large $\NN=(4,4)$. A similar enhancement for $k=2$ is possible but even less obvious at the moment 
(see comments in section \ref{lift}).

We conclude in section \ref{outlook} with a brief discussion of interesting aspects of the proposed action 
and its implications in M-theory. Open problems that are worth pursuing further are also discussed in this
section.

\section{M2-M5 from the M-theory lift of a type IIB setup}
\label{lift}

\subsection{Type IIB setup}
\label{Blift}

Our starting point is the following Hanany-Witten setup in type IIB string theory that realizes at low energies 
the ABJM model \cite{Aharony:2008ug} on a space with a boundary
\beq
\label{liftaa}
	\begin{array}{r c c c c c c c c c c c}
	N ~ D3 & ~:~ & 0  & 1 & 2_+ &  &  &  & 6_+ &  &  & 
	\\
	N ~ D3' & ~:~ & 0  & 1 & 2_+ &  &  &  & 6_- &  &  & 
	\\
	1 ~ NS5  & ~:~ & 0  & 1 & 2 & 3 & 4 & 5 &  &  &  & 
	\\
	1 ~(1,k)5 & ~:~ & 0  & 1 & 2 &  \left[ {3 \atop 7} \right]_{\theta} &  \left[ {4 \atop 8} \right]_{\theta} 
	& \left[ {5 \atop 9} \right]_{\theta} &  &  &  & 
	\\
	M ~ D5  & ~:~ & 0  & 1 &  &  &  &  & 6 & 7 & 8 & 9
	\end{array}
\eeq

In this setup an NS5-brane and a $(1,k)$5-brane bound state\footnote{We will be using conventions where 
$(p, q)5$ refers to a fivebrane bound state with $p$ units of NS5-brane charge and $q$
units of D5-brane charge. Moreover, without loss of generality we will henceforth assume that 
$k > 0$. The notation $\left[ {a \atop b} \right]_\theta$ denotes that a brane is oriented along the direction
$\cos\theta\, x^a +\sin\theta\, x^b$ in the $(x^a, x^b)$ plane.} 
are located at antipodal points on the $S^1$ direction $x^6 \in [-\pi,\pi)$. 
The angle $\theta$ is fixed by supersymmetry in terms of the complex axion-dilaton coupling $\tau$
\beq
\label{liftab}
\theta = {\rm arg}(\tau) - {\rm arg}(k+\tau)~, ~~ \tau =\frac{i}{g_s}+\chi
\eeq
where $g_s$ is the string coupling constant and $\chi$ the value of the axion field (that we set to zero).
Two stacks of D3-branes are suspended between the NS5 and $(1,k)5$-branes 
along the directions $(0126)$: $N$ D3 branes wrap 
the semi-circle $x^6\in (0,\pi)$, and $N$ D3$'$ branes wrap the semi-circle $x^6 \in (-\pi, 0)$. 
The setup of D3-NS5-$(1,k)5$ branes, with the D3-branes stretching infinitely across the whole $(012)$ plane
and $M=0$ D5-branes, 
is the original configuration of Ref.\ \cite{Aharony:2008ug} that formulated the low-energy theory on $N$
M2-branes probing ${\mathbb C}^4/{\mathbb Z}_k$ as a $U(N)_k \times U(N)_{-k}$ Chern-Simons-matter theory.
For $k>2$ this theory is an $\NN=6$ three-dimensional gauge theory. For $k=1,2$ there is an infrared enhancement
of supersymmetry to $\NN=8$ \cite{Aharony:2008ug,Bashkirov:2010kz}.

Compared to Ref.\ \cite{Aharony:2008ug},
the setup \eqref{liftaa} introduces an additional stack of $M$ D5-branes
(last line in \eqref{liftaa}) that intersect the $N$ pairs of D3-branes on 
a two-dimensional boundary along the plane $(01)$. The semi-infinite D3-branes stretch on the half-line 
$x^2>0$ and end on the D5-branes at $x^2=0$ (hence the notation $2_+$ in the list of 
the configuration \eqref{liftaa}). From the low-energy point of view, the D5-branes 
introduce a boundary on the three-dimensional Chern-Simons-matter theory 
that resides on the D3-branes. One can verify by explicit computation (see e.g.\ appendix A 
of Ref.\ \cite{Armoni:2015jsa} for a related discussion) that the brane setup 
\eqref{liftaa} preserves 3 real supersymmetries ---2 left-moving and 1 right-moving. Hence this is a non-chiral 
half-BPS boundary. At low-energies the global symmetries of the M-theory lift (to be discussed momentarily) suggest
the infrared enhancement of supersymmetry to $\NN=(4,2)$ in two dimensions. In the special case where 
$k=1$ they suggest a further enhancement to large $\NN=(4,4)$.

As an aside remark, it is useful to note here, for later purposes, the following fact. 
Rotating the $(1,k)5$-brane in \eqref{liftaa} along the more general
orientation $\left( \left[ {3 \atop 7} \right]_{\psi}~ \left[ {4 \atop 8} \right]_{\psi}~ \left[ {5 \atop 9} \right]_{\theta} \right)$ 
further reduces the explicit supersymmetry from 3 real supersymmetries to 2 real supersymmetries when 
$\psi \neq \theta$. Namely, changing $\psi$ reduces $\NN=(2,1) \to \NN=(1,1)$ in two dimensions. 

In section \ref{qft} we consider the low-energy field theory at the D3-D5 intersection  
following a recent similar discussion of open string dynamics in \cite{Armoni:2015jsa}. 
In the rest of this section we elaborate further on the M-theory lift of the setup \eqref{liftaa} 
and its relation to the orthogonal M2-M5 intersection which is the system of main interest in this paper.

\subsection{M-theory lift}
\label{Mlift}

Repeating the steps of the U-duality transformation in \cite{Aharony:2008ug} we first perform a T-duality 
transformation along the direction 6. This results to a type IIA brane configuration on a space with a T-dual 
$S^1$ direction $\tilde 6$:
\beq
\label{liftac}
	\begin{array}{r c c c c c c c c c c c}
	N ~ D2_+ & ~:~ & 0  & 1 & 2_+ &  &  &  &  &  &  & 
	\\
	1 ~ KK_{\tilde 6}  & ~:~ & 0  & 1 & 2 & 3 & 4 & 5 &  &  &  & 
	\\
	1 ~ (KK_{\tilde 6}-k\, D6) & ~:~ & 0  & 1 & 2 &  \left[ {3 \atop 7} \right]_{\theta} &  \left[ {4 \atop 8} \right]_{\theta} 
	& \left[ {5 \atop 9} \right]_{\theta} &  &  &  & 
	\\
	M ~ D4  & ~:~ & 0  & 1 &  &  &  &  &  & 7 & 8 & 9
	\end{array}
\eeq
The notation $KK_{\tilde 6}$ refers to a Kaluza-Klein (KK) monopole associated with the dual circle $\tilde 6$.
Similarly $(KK_{\tilde 6}-k\, D6)$ refers to a KK monopole with $k$ units of D6-brane flux.

Next we lift to M-theory by adding the 11-th (M-theory) direction $x^{10}$. The $N$ D2-branes ending on $M$ 
D4-branes become $N$ M2-branes ending on $M$ M5-branes. The KK monopole $KK_{\tilde 6}$ remains a
KK monopole associated with $\tilde 6$ and the $(KK_{\tilde 6}-k\, D6)$ bound state becomes a KK monopole
associated with a linear combination of the circles $\tilde 6$ and $10$. At the intersection of the two KK
monopoles the eight-dimensional space transverse to the plane $(012)$ becomes the orbifold 
${\mathbb C}^4/{\mathbb Z}_k$ \cite{Aharony:2008ug}. 

To summarize, after the above U-duality transformation we obtain the orthogonal M2-M5 intersection 
\beq
\label{liftad}
	\begin{array}{r c c c c c c c c c c c c}
	N ~ M2_+ & ~:~ & 0  & 1 & 2_+ &  &  &  &  &  &  &  &
	\\
	M ~ M5  & ~:~ & 0  & 1 &  &  &  &  &  & 7 & 8 & 9 & 10
	\end{array}
\eeq
probing the ${\mathbb C}^4/{\mathbb Z}_k$ singularity in the $(3456789(10))$ directions.
The $k=1$ case reduces to the familiar M2-M5 intersection in flat space.

As explained in appendix B of Ref.\ \cite{Aharony:2008ug} the metric of the transverse eight-dimensional space
takes the form of a toric hyperk\"ahler manifold with a diagonal $2\times 2$ matrix of $U$-functions in the coordinates
\beq
\label{liftaea}
\vec x'_1 = (x^7, x^8, x^9)~, ~~ \varphi'_1 = x^{\tilde 6} - \frac{1}{k} x^{10}
\eeq 
and
\beq
\label{liftaeb}
\vec x'_2 = (x^7+k x^3, x^8+k x^4, x^9+k x^5)~, ~~ \varphi'_2 = \frac{1}{k} x^{10}
~.
\eeq
The coordinates $(\varphi'_1,\varphi'_2)$ have periodicity $2\pi$ plus the orbifold identification
\beq
\label{liftaec}
(\varphi'_1,\varphi'_2) \sim (\varphi'_1,\varphi'_2) +\left( -\frac{1}{k}, \frac{1}{k} \right)
~.
\eeq
In the absence of the M5-branes the overall symmetry of the transverse space is $SO(6) \times SO(2)$.
$SO(6)$ is associated with transformations in the $(345789)$ directions and $SO(2)$ with translations of $x^{10}$,
i.e.\ with translations $(\varphi_1',\varphi'_2) \to (\varphi_1'-\varphi,\varphi'_2+\varphi)$.

In the presence of the M5-branes the $SO(6)$ in the $(345789)$ directions breaks to $SO(3) \times SO(3)$
transformations that are either fully parallel to the M5-brane worldvolume or fully orthogonal. Since 
$SO(3) \simeq SU(2)$ and $SU(2)\times SU(2) \simeq SO(4)$, the total symmetry of the M2-M5 configuration
in the presence of the orbifold, for $k>2$, is $SO(4) \times SO(2)$. This is an R-symmetry for the 
two-dimensional theory at the M2-M5 intersection. Its presence suggests
that the infrared theory at the intersection exhibits $\NN=(4,2)$ supersymmetry. 

For $k=1$ the symmetry of the transverse ${\mathbb R}^8$ is $SO(4)\times SO(4)$ from the separate rotation 
symmetries of the two orthogonal ${\mathbb R}^4$'s in ${\mathbb R}^8$. 
This symmetry is the expected R-symmetry group of a 2d CFT with 
large $\NN=(4,4)$ superconformal algebra. 

The case with $k=2$ is potentially more interesting. In the absence of the M5-branes arguments were
given in \cite{Aharony:2008ug} for the quantum mechanical enhancement of the R-symmetry group in the 
three-dimensional Chern-Simons-matter theory from $SO(6)$ to $SO(8)$.
In our setup a stack of M5-branes intersects the ${\mathbb Z}_2$ singularity. If the non-abelian interactions of the 
M5-brane theory exhibit the same global symmetry enhancement one would expect an $SO(4)\times SO(4)$
symmetry for the M2-M5 intersection also at $k=2$. It is currently unclear to us if this enhancement actually takes
place.

\section{ABJM on a space with boundary from open string theory}
\label{qft}

In this section we focus on the open string theory dynamics of the type IIB setup \eqref{liftaa}. Following the
discussion of the recent paper \cite{Armoni:2015jsa} we propose a specific action for the 3d-2d bulk-boundary
dynamics at the D3-D5 intersection.

\subsection{3d bulk}

The 3d bulk theory, which arises as the IR effective field theory description of the open string dynamics on the
D3-branes in the setup \eqref{liftaa}, is the $\NN=6$ $U(N)_k \times U(N)_{-k}$ ABJM theory. It is formulated
most conveniently as an $\NN=2$ theory with appropriate matter representations.
The Lagrangian for the $\NN=2$ vector multiplet consists of the $\NN=2$ Chern-Simons (CS) theory
at level $k$, and the $\NN=2$ CS theory at level $-k$. The gauge group of both CS theories is $U(N)$.
To distinguish between the two gauge groups we will denote them as $U(N)_+$ (with CS level $+k$), and
$U(N)_-$ (with CS level $-k$).

The matter content of the theory consists of 2 chiral superfields $A^a$ ($a=1,2$) in the bifundamental 
representation of $U(N)_+ \times U(N)_-$ and 2 chiral superfields $B_{a}$ ($a=1,2$) in the anti-bifundamental
representation. (The complex conjugate anti-chiral superfields will be denoted with a bar.) It is convenient,
and most appropriate from the point of view of the brane configuration \eqref{liftaa}, to include two massive 
$\NN=2$ chiral superfields $\phi_\pm$ with superpotential
\beq
\label{qftaa}
W= \frac{k}{8\pi} \Tr \left[ \phi^2_+ - \phi^2_- \right] +\Tr \left[ B_a \phi_+ A^a \right] 
+ \Tr \left[ A^a \phi_- B_a \right]
~.
\eeq
Integrating out the massive superfields sets
\beq
\label{qftaaaA}
\phi_+ = - \frac{4\pi}{k} A^a B_a ~, ~~ 
\phi_- = \frac{4\pi}{k} B_a A^a 
\eeq
and leads in the deep IR to the quartic superpotential 
\beq
\label{qftaaa}
W= \frac{4\pi}{k} \Tr \left[ A^1 B_1 A^2 B_2 - A^1 B_2 A^2 B_1 \right]
= \frac{2\pi}{k} \varepsilon_{ab} \varepsilon^{\dot a \dot b} 
\Tr \left[ A^a B_{\dot a} A^b B_{\dot b} \right]
\eeq
which is responsible for the supersymmetry enhancement to $\NN=6$. $\varepsilon_{ab}$ is the totally anti-symmetric
symbol; we use $\varepsilon_{12}=1$.

It is useful to highlight the following points regarding the $\NN=2$ formulation of the ABJM theory:
\begin{itemize}
\item[$(i)$] A general mass $m\neq \frac{k}{8\pi}$ in \eqref{qftaa} corresponds in the brane setup \eqref{liftaa} to a 
general angle $\psi \neq \theta$ for the $(1,k)5$-brane oriented along 
$\left( \left[ {3 \atop 7} \right]_{\psi}~ \left[ {4 \atop 8} \right]_{\psi}~ \left[ {5 \atop 9} \right]_{\theta} \right)$, 
\cite{Kitao:1998mf,Bergman:1999na}. 
As we pointed out near the end of subsection \ref{Blift}, and is visible from \eqref{qftaa}, for $\psi\neq \theta$ 
and $m\neq \frac{k}{8\pi}$ the explicit supersymmetry of the 3d bulk theory is $\NN=2$ (and therefore 2d $\NN=(1,1)$
on a half-BPS boundary). Nevertheless, it was shown perturbatively in \cite{Gaiotto:2007qi} in the large $k$ limit that the 
$\NN=6$ fixed point is an attractor of the RG flow in the 3d Chern-Simons-matter theory,
so different values of $m$ in the bare action do not affect the IR physics crucially in the bulk. It is natural to expect
a similar effect for all values of $k$. 
For technical reasons that will become clear momentarily, it will be useful to work 
with a general mass $m$ in the bulk superpotential
\beq
\label{qftaab}
W= m \Tr \left[ \phi^2_+ - \phi^2_- \right] +\Tr \left[ B_a \phi_+ A^a \right] 
+ \Tr \left[ A^a \phi_- B_a \right]
~.
\eeq

\item[$(ii)$] The bare $\NN=2$ supersymmetric action with superpotential \eqref{qftaa} does not exhibit the 
$\NN=3$ supersymmetry automatically in the non-abelian case unless some of the auxiliary fields in the 
$\NN=2$ supersymmetric multiplets are integrated out. Hence, we would not expect to see the full
$\NN=(2,1)$ supersymmetry on the half-BPS 2d boundary in the UV in the above language in a fully 
$\NN=2$ super-gauge invariant formulation. Note that the abelian case does not exhibit this issue. 

\item[$(iii)$] Finally, working with explicit $\NN=2$ supersymmetry in the bulk allows us to circumvent an important
technical issue that has to do with the effects of the boundary. It is well-known that boundaries break the 
super-gauge invariance of supersymmetric gauge theories. Therefore, the passage to a preferable gauge 
may be inappropriate in the presence of a boundary. As a result,
a proper treatement of boundaries typically requires a formulation with full off-shell supersymmetry. 
For example, in the case of the $\NN=6$ Chern-Simons-matter theories of interest this would require the use 
of an explicit $\NN=6$ formalism, which is a rather complicated task. 

We circumvent this problem by 
formulating the half-BPS boundary and the corresponding bulk-boundary interactions in the bare 
$\NN=2$ Lagrangian with superpotential \eqref{qftaab}. 
Then by tuning the bare mass $m$ to the $\NN=3$ point $m=\frac{k}{8\pi}$, or by just allowing the
renormalization group to flow to the $\NN=6$ fixed point in the bulk, 
we postulate that our half-BPS boundary flows accordingly from a UV $\NN=(1,1)$ point   
to the desired IR point with $\NN=(4,2)$ (or $\NN=(4,4)$) supersymmetry. 
We provide favorable evidence for this conjecture using the available information 
from string theory and by checking explicitly that the postulated action has the expected 
global symmetries.
 
\end{itemize}

\subsubsection*{\it 3d bulk action}

In the brane setup \eqref{liftaa} there is a boundary for the 3d theory at $x^2=0$. Accordingly, we will 
formulate the ABJM theory on the half-plane at $x^2>0$. We use the $\NN=2$ superspace formalism 
and the set of conventions summarized in \cite{Armoni:2015jsa}.\footnote{The superspace coordinates are
$(x^\mu,\vartheta_\alpha)$ with spacetime indices $\mu=0,1,2$ and spinor indices $\alpha=\pm$. The
odd Grassmann variables $\vartheta_\alpha$ are complex: 
$\vartheta_\alpha = \frac{1}{\sqrt 2}\left(\theta_{1\alpha}+i \theta_{2\alpha}\right)$. $\theta_{s\alpha}$ $(s=1,2)$ are
real Grassmann odd variables in $\NN=1$ superspace. We follow the $\NN=1$ superspace conventions in 
Ref.\ \cite{Gates:1983nr}.} In these conventions the content of the $\NN=2$ 
vector multiplet is summarized in an $\NN=2$ vector superfield $V$ that contains the three-dimensional 
gauge field $A_\mu$, several auxiliary scalars and their supersymmetric partners. As we mentioned already, 
in ABJM there are two vector multiplets that we call $V_+$, $V_-$, which appear in the CS actions with level 
$+k$ and $-k$ respectively.
The fully covariant formulation of the $\NN=2$ CS theory is four-dimensional \cite{Ivanov:1991fn}. In our context
\bea
\label{qftab}
\SS_{CS}[V_+,V_-] &=&
-\frac{k}{2\pi} \int_0^1 ds \int d^3 x \int d^4 \vartheta \, \Tr \left[ V_+ \bar {\boldsymbol D}^\alpha \left( e^{sV_+} 
{\boldsymbol D}_\alpha e^{-sV_+} \right)\right]
\nonumber\\
&+& \frac{k}{2\pi} \int_0^1 ds \int d^3 x \int d^4 \vartheta \, \Tr \left[ V_- \bar {\boldsymbol D}^\alpha \left( e^{sV_-} 
{\boldsymbol D}_\alpha e^{-sV_-} \right)\right]
~.
\eea
${\boldsymbol D}_\alpha$ is the $\NN=2$ superspace covariant derivative
\beq
\label{qftac}
{\boldsymbol D}_\alpha = \partial_\alpha + \left( \gamma^\mu \bar \vartheta \right)_\alpha \partial_\mu~, ~~
\bar {\boldsymbol D}_\alpha = \bar \partial_\alpha + \left( \gamma^\mu \vartheta \right)_\alpha \partial_\mu
~.
\eeq

The matter sector interactions include the superpotential interactions \eqref{qftaab} (or \eqref{qftaa}
for the specific orientations in \eqref{liftaa})
\beq
\label{qftad}
\SS_W[\phi_\pm, A,B] = \int d^3 x \, d^2\vartheta \, W + {\rm c.c.}
~,
\eeq
and the K\"ahler interactions that provide the kinetic terms. For simplicity, we will consider here canonical
K\"ahler interactions. Note however, that, unlike the superpotential interactions, the 
K\"ahler interactions receive quantum corrections. Accordingly, the boundary interactions that will be formulated
shortly have to be adjusted suitably to take into account these quantum corrections in order to preserve the
desired amount of supersymmetry. This can be performed straightforwardly with the prescription
that will be described in a moment. 
The canonical K\"ahler interactions that we consider here are
\bea
\SS_{K}[\phi_\pm, A,B,V_\pm] &=& \int d^3 x \int d^4\vartheta \, \Tr
\Big[ 
\bar \phi_+ e^{V_+} \phi_+ 
+ \bar\phi_- e^{V_-} \phi_-
+ \bar A_a e^{V_+} A^a e^{- V_-} 
+ \bar B^{a} e^{ V_-} B_{a} e^{- V_+} 
\Big]
~.
\nonumber\\
\eea

In summary, the total bulk action is
\beq
\label{qftae}
\SS_{bulk} = \SS_{CS}[V_+,V_-] + \SS_K[\phi_\pm, A, B, V_\pm] +\SS_W [\phi_\pm,A,B]
~.
\eeq

\subsection{2d boundary}

\begin{figure}[t!]
   \centering
    \includegraphics{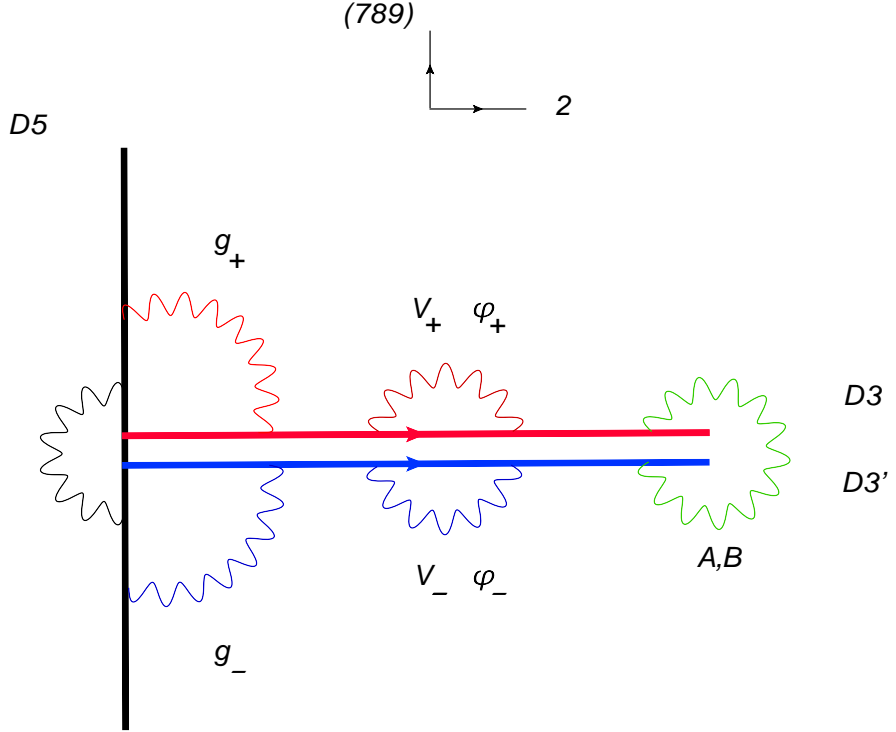}
    {\bf \caption{\rm Two stacks of $N$ D3-branes (wrapping different halves of the $6$-circle) end from the right 
    $(x^2>0)$ on $M$ D5-branes within the brane configuration \eqref{liftaa}. We have isolated the D3-D5 part
    of the intersection in the directions (2789) leaving the NS5, $(1,k)5$ part of the setup implicit.
    }
   \label{D3D5}
   }
\end{figure}

The boundary theory of a Chern-Simons-matter theory is not unique. In previous explorations of the subject
\cite{Berman:2009kj,Chu:2009ms,Faizal:2011cd} boundary interactions were formulated with 
two main guiding criteria: the restoration of the appropriate amount of gauge invariance and supersymmetry.
Even with these criteria there is still considerable freedom on the choice of boundary degrees of freedom and 
boundary/bulk-boundary interactions. As a result, a well-motivated specific proposal for the boundary theory of $N$ 
M2-branes ending on $M$ M5-branes has not been possible so far. 

In the context of the brane setup \eqref{liftaa} we find ourselves in a much better situation. Following the recent
discussion in \cite{Armoni:2015jsa} we can now use the open string theory of the type IIB Hanany-Witten setup as a 
concrete guide towards a boundary action. Different sectors of the open string theory at the D3-D5 intersection
are summarized Fig.\ \ref{D3D5}. Besides the 3d bulk fields $V_\pm$, $\phi_\pm$, $A^a$, $B_{a}$ from 3-3,
$3'$-$3'$, or 3-$3'$ open strings there are also $g_+$ fields from 3-5 (red) strings and $g_-$ fields from 5-$3'$ 
(blue) strings. Both are 2d $\NN=(1,1)$ superfields; $g_+$ is in the bifundamental representation of 
$U(N)_+ \times U(M)$ and $g_-$ in the bifundamental of $U(M) \times U(N)_-$. There are also fields
from 5-5 strings (black color) which will be ignored since their dynamics is irrelevant at low energies.
The group representations in which different supermultiplets belong are summarized in Table \ref{table:reps}.

\begin{table}[htbp]
   \centering
      \begin{tabular}{@{} lc c r @{}} 
      \toprule
            Superfield    & $U(N)_-$ & $U(N)_+$ & $U(M)$ \\
      \midrule
      $V_+$      &  {\bf 1} &  adjoint & {\bf 1}\\
      $V_-$          & adjoint     &  {\bf 1} & {\bf 1}\\
      $\phi_+$       & {\bf 1}  & adjoint  & {\bf 1}\\
      $\phi_-$       & adjoint  & {\bf 1} & {\bf 1} \\
      $A^a$ &  $\Box$   &  $\overline{\Box}$ & {\bf 1} \\
      $B_a$ &  $\overline{\Box}$ & $\Box$ & {\bf 1} \\
      $g_+$  & {\bf 1} & $\overline{\Box}$ & $\Box$ \\
      $g_-$  &  $\Box$  & {\bf 1}  & $\overline{\Box}$ \\
      \bottomrule
   \end{tabular}
   {\bf \caption{\rm A summary of group representations.}
   \label{table:reps}}
\end{table}

The first step in the construction of a boundary action involves the addition of suitable boundary interactions
that restore the desired amount of supersymmetry. In the case at hand we have explicit $\NN=2$ supersymmetry
in the bulk and want to restore $\NN=(1,1)$ supersymmety on the boundary. Applying the prescription of Ref.\
\cite{Belyaev:2008xk} to a general $\NN=2$ action (expressed conveniently in $\NN=1$ superspace language)
\beq
\label{qftba}
\SS = \int d^3x \, d^2 \theta_1 \, d^2\theta_2 \, \LL(x^\mu,\theta_1,\theta_2)
\eeq
we obtain the action
\beq
\label{qftbb}
\SS^{(1,1)} = \int d^3x \Big\{ d^2 \theta_1 \, d^2\theta_2 \, \LL 
-d^2 \theta_1 \, \partial_2 \LL \Big |_{\theta_2=0} 
+ d^2\theta_2 \, \partial_2 \LL \Big |_{\theta_1=0}
- \partial_2 \partial_2 \LL \Big |_{\theta_1=\theta_2=0} \Big\}
\eeq
that preserves the supersymmetries generated by $(Q_{1+},Q_{2-})$. In our case, $\SS = \SS_{bulk}$
(see eq.\ \eqref{qftae}). We will denote this boundary-corrected version of $\SS_{bulk}$
\beq
\label{qftbba}
\SS^{(1,1)}_{bulk} = \SS^{(1,1)}_{CS} + \SS^{(1,1)}_K + \SS^{(1,1)}_W
~.
\eeq

The next step involves the incorporation of the boundary multiplets $g_\pm$ 
in a manner that restores the broken $U(N)\times U(N)$ gauge invariance at the boundary. 
Following \cite{Armoni:2015jsa} we extend the definition of $g_\pm$ in the bulk as 3d $\NN=2$ superfields
(denote them ${\boldsymbol g}_\pm$), 
and define the $U(M)$ $\NN=2$ vector superfields ${\boldsymbol V}^{g_\pm}_\pm$ 
\beq
\label{qftbc}
e^{{\boldsymbol V}^{g_+}_+} = \tilde {\boldsymbol g}_+ e^{V_+} {\boldsymbol g}_+
~, ~~
e^{{\boldsymbol V}^{g_-}_-} = \tilde {\bar {\boldsymbol g}}_- e^{V_-}   {\bar {\boldsymbol g}}_-
~.
\eeq
We are using the notation
\beq
\label{qftbca}
\tilde {\boldsymbol g} = \bar {\boldsymbol g} \left( {\boldsymbol g} \bar {\boldsymbol g} \right)^{-1}
\eeq
that has the useful property
\beq
\label{qftcb}
{\boldsymbol g} \tilde {\boldsymbol g} = {\bold 1}_{N\times N}
~.
\eeq
$\bar {\boldsymbol g}$ is the Hermitian conjugate of ${\boldsymbol g}$.
The proposed boundary interactions for the $g_\pm$ bifundamentals are \cite{Armoni:2015jsa}
\beq
\label{qftbd}
\SS_{bdy}^{(gauge)} = 
\SS_{kin}^{(1,1)}\left[ g_+,V_+ \right] + \SS_{kin}^{(1,1)}\left[ g_-,V_- \right] 
+\SS^{(1,1)}_{CS} \left[ {\boldsymbol V}_+^{g_+}, {\boldsymbol V}_-^{g_-} \right] 
- \SS^{(1,1)}_{CS} \left[ V_+, V_- \right] 
~.
\eeq
$\SS^{(1,1)}_{kin}$ provides $\NN=(1,1)$ supersymmetric, gauge-invariant kinetic terms. 
In more explicit form
\bea
\label{qftbe}
&&\SS^{(1,1)}_{kin} \left[ g_+,V_+ \right] + \SS_{kin}^{(1,1)}\left[ g_-,V_- \right] 
\\
&&= - \frac{k}{2\pi} \int d^2 x \int d\theta_{1+} d\theta_{2-} \,
\Big\{ \left( {\bar g}_+ \hat {\boldsymbol \nabla}^{(+)}_- g_+ \right) 
\left(  {\bar g}_+ \hat {\boldsymbol \nabla}^{(+)}_+   g_+ \right)
+  \left(  g_- \hat {\boldsymbol \nabla}^{(-)}_-  {\bar g}_- \right) 
\left( g_- \hat {\boldsymbol \nabla}^{(-)}_+   {\bar g}_- \right) \Big\}
\nonumber
\eea
where the light-cone coordinates $x^\pm = x^0 \pm x^1$ were used, 
and $\hat {\boldsymbol \nabla}^{(\pm)}_\alpha$ 
are the boundary versions of the chiral $\NN=2$ super-gauge-covariant derivatives 
\beq
\label{qftbg}
{\boldsymbol \nabla}_\alpha^{(\pm)} = e^{-V_\pm} {\boldsymbol D}_\alpha e^{V_\pm}
~.
\eeq

The last two terms on the rhs of equation \eqref{qftbd} are a difference of two four-dimensional 
actions. This difference is a total derivative term that contributes Wess-Zumino-like interactions 
for $g_\pm$ supported only on the two-dimensional boundary. To obtain this result one has to 
employ the property \eqref{qftcb}. 

So far the total bulk-boundary action is
\beq
\label{qftbi}
\SS_{bulk}^{(1,1)} + \SS_{bdy}^{(gauge)} 
= \SS_{kin}^{(1,1)}\left[ g_+,V_+ \right] + \SS_{kin}^{(1,1)}\left[ g_-,V_- \right] 
+\SS^{(1,1)}_{CS} \left[ {\boldsymbol V}_+^{g_+}, {\boldsymbol V}_-^{g_-} \right] 
+\SS^{(1,1)}_K + \SS^{(1,1)}_W
~.
\eeq
There are explicit couplings of the boundary degrees of freedom $g_\pm$ with the vector multiplets $V_\pm$, but 
no couplings with the other bulk superfields, $\phi_\pm$, $A^a$, $B_a$. From the string/M-theory discussion in 
section \ref{lift} we recall that the boundary is expected to break the bulk $SO(6)\times SO(2)$ R-symmetry to 
$SO(4) \times SO(2)$ and the action \eqref{qftbi} does not have this property. This is already an indication that
the open string theory of the D3-D5 intersection in configuration \eqref{liftaa} involves additional boundary interactions.

From the open string theory of the configuration represented in Fig.\ \ref{D3D5} it is indeed clear that there is a cubic
interaction on the two-dimensional boundary of the form
\beq
\label{qftbj}
\SS_{bdy}^{(matter)} = 
\int d^2 x \int d\theta_{1+} d\theta_{2-} \, \Tr
\left[
g_- \left( \hat \phi_- +\hat{\bar\phi}_-\right) \bar g_-
+\bar g_+ \left( \hat \phi_+ + \hat{\bar \phi}_+ \right) g_+
\right]
~.
\eeq
$\hat \phi_\pm$ denotes the $\NN=(1,1)$ projection of the bulk superfields $\phi_\pm$ on the two-dimensional 
boundary. For a succinct summary of boundary projections of superfields see appendix 3 of Ref.\ \cite{Berman:2009kj}.
More precisely, in the particular context of eq.\ \eqref{qftbj} by $\hat \phi_\pm$ we refer to the projection
\beq
\label{qftbja}
\hat \phi_\pm (\theta_{1+},\theta_{2-})= \widetilde{\widehat {\Big( e^{V_\pm} \, \phi_\pm \, e^{-V_\pm} \Big)}}
\eeq
where the notation \,$\tilde \hat{}$\, refers to the notation of eq.\ (198) in \cite{Berman:2009kj}. 
The $\NN=2$ vector field exponents have been inserted to gauge-covariantize the derivatives normal to the boundary
that appear in the \,$\tilde \hat{}$\, projection. Notice that this cubic interaction has exactly the same form with a 
corresponding bulk-boundary interaction that appears in the field theory of the flat-space D3-D5 intersection 
\cite{Karch:2001cw,DeWolfe:2001pq}. Although the physics of the flat-space D3-D5 intersection (without the additional
5-branes of the HW setup that we consider) is considerably different from the physics of our setup the uniqueness of 
this cubic interaction in \cite{DeWolfe:2001pq} and its crucial role in the expected supersymmetry enhancement in 
that context, gives some confidence that \eqref{qftbja} is the only extra interaction that we need to include at
low energies.

As a further more direct check, we notice that the expected symmetries, e.g.\ invariance under the ${\mathbb Z}_k$ 
transformations
\beq
\label{qftbk}
A^a \to e^{2\pi i/k} A^a~, ~~ B_a \to e^{-2\pi i/k} B_a ~, ~~ g_\pm \to e^{\mp 2\pi i/k} g_\pm
~,
\eeq
do not allow cubic $U(M)$-invariant interactions between $g_\pm$ and the bi-fundamental fields $A^a$, $B_a$.

Moreover, the boundary interaction \eqref{qftbj} implements the breaking of the R-symmetry
\beq
\label{qftbl}
SO(6) \times SO(2) \to SO(4) \times SO(2)
\eeq
that was anticipated from string/M-theory in section \ref{lift}. This can be verified explicitly in the following way. 
In the three-dimensional 
bulk the UV action with the massive $\phi_\pm$ fields exhibits an $SU(2)_{diag}$ symmetry that rotates simultaneously 
the bottom components of the $A^a$ and $B_a$ superfields. In the IR the quartic superpotential \eqref{qftaaa} 
enhances this symmetry to $SU(2)_A \times SU(2)_B$ that rotates independently the fields $A^a$, $B_a$. These
two $SU(2)$'s together with an independent $SU(2)_R$ symmetry that rotates the fields $(A^1, B^*_1)$
combine to the $SO(6)$ of the bulk action. On the boundary the interaction \eqref{qftbj} respects only the diagonal
$SU(2)_{diag}$ symmetry of $SU(2)_A \times SU(2)_B$ and does not allow it to enhance in the IR. Hence, in the 
infrared we expect the theory to exhibit the overall global symmetry $SU(2)_{diag} \times SU(2)_R \times SO(2)
\sim SO(4) \times SO(2)$. 

Notice, that by integrating out the massive $\phi_\pm$ fields, setting $m=\frac{k}{8\pi}$, and using the 
identification \eqref{qftaaaA}, the boundary interaction \eqref{qftbj} turns into the quartic interaction
\beq
\label{qftbm}
\SS_{bdy}^{(matter)} = \frac{4\pi}{k}
\int d^2 x \int d\theta_{1+} d\theta_{2-} \, \Tr_{U(M)}
\Big[
g_- \left( \hat B_a \hat A^a +\hat {\bar A}^{a} \hat {\bar B}_a\right) \bar g_- 
- \bar g_+ \left( \hat A^a \hat B_a + \hat {\bar B}_a \hat {\bar A}^{a} \right) g_+ 
\Big]
~.
\eeq
Observe that further interactions of the components of the $A^a$, $B_a$ superfields will be induced 
on the boundary by this integrating out procedure from the boundary terms included in 
$\SS_K^{(1,1)}+\SS_W^{(1,1)}$ according to the prescription \eqref{qftbb}.

\subsection{Summary of the proposed bulk-boundary action}

Collecting all the interactions in favor of which we argued above, we propose that the infrared limit of the bare 
bulk-boundary action
\bea
\label{qftca}
&&\SS_{proposed}[g_\pm, V_\pm, \phi_\pm, A, B] = \SS_{bulk}^{(1,1)} + \SS_{bdy}^{(gauge)} + \SS_{bdy}^{(matter)}
\nonumber\\
&&= \SS_{kin}^{(1,1)}\left[ g_+,V_+ \right] + \SS_{kin}^{(1,1)}\left[ g_-,V_- \right] 
+\SS^{(1,1)}_{CS} \left[ {\boldsymbol V}_+^{g_+}, {\boldsymbol V}_-^{g_-} \right] 
+\SS^{(1,1)}_K + \SS^{(1,1)}_W
+\SS_{bdy}^{(matter)}
\eea
describes the low-energy theory at the M2-M5 intersection \eqref{liftad}.
All the terms that appear in \eqref{qftca} were defined previously in the main text. 
We will not attempt to write out this action in components. In appendix \ref{oneM2} we present a more explicit
form of the interactions in the case of a single M2-brane ending on an arbitrary number of M5-branes.  
The part that is hardest to expand in components is the non-abelian $\NN=2$ Chern-Simons action \eqref{qftab}, which is 
written as a four-dimensional integral. There is such a non-abelian term, for general $M$, even in the abelian case
of a single M2-brane, $N=1$.

\section{Outlook}
\label{outlook}

In this paper we conjectured a specific action for the infrared theory in the M2-M5 intersection \eqref{liftad}
with explicit $\NN=2$ supersymmetry in the bulk and explicit $\NN=(1,1)$ supersymmetry on the two-dimensional 
boundary. The boundary degrees of freedom and their interactions were deduced in large part from the open string theory 
of the type IIB Hanany-Witten configuration \eqref{liftaa}. Some evidence from the proposed interactions follows
from the consistency of the constructions in \cite{Armoni:2015jsa}. We verified the expected global symmetries, and 
postulated that this action should exhibit the required $SO(4)\times SO(2)$ R-symmetry in the deep infrared. 
Accordingly we conjectured the enhancement of the boundary supersymmetry to $\NN=(4,2)$ for $k>2$. 

It would be very useful to find further checks of this preliminary proposal and eventually prove conclusively 
that it is the correct infrared description of M2-M5 physics. 
In this context, it would be interesting to explore the relation of this work with the Basu-Harvey equations 
\cite{Basu:2004ed}. It would also be interesting to explore relations with the work \cite{Gomis:2014eya} upon 
compactification. In that respect, notice that the 2d boundary theories in \cite{Gomis:2014eya} also
include a pair of bi-fundamentals, which are analogous to our $g_\pm$.

Having a UV bare action is a first step towards the analysis of the quantum properties of the M2-M5
system. Generically this system is strongly coupled, but the introduction of the CS level $k$ opens the possibility
to go in weak coupling regimes. These are roughly regimes where the ratio $N/k$ is small. It would be interesting
to explore these regimes with perturbative techniques. 

One of the issues that would be worth understanding better is whether the 2d boundary theory has a well-defined
decoupling limit with a conserved 2d stress-energy tensor. One can then ask about the central charge of the
boundary theory, and how it depends on the three parameters $N, M, k$. Our UV action introduces the
massless boundary degrees of freedom $g_\pm$ which belong in the bi-fundamental representation 
of $U(N) \times U(M)$. Hence, their number scales as $N M$ in agreement with the anomaly considerations 
of Ref.\ \cite{Berman:2004ew}.
In the IR the corresponding central charge can exhibit different scalings, similar to the reduction observed in the 
ABJM theory, where the $N^2$ UV scaling of the massless degrees of freedom reduces in the IR to 
the familiar $N^{3/2}$. It would be very interesting to see if the action that we propose has a consistent
't-Hooft like limit with $N,M\gg 1$ and the ratio $\lambda = M^2/N$ fixed, and if the central charge of the 
boundary theory scales in the large $\lambda$-limit as predicted by the blackfold supergravity analysis \eqref{setupab}.

\subsection*{Acknowledgments}

I would like to thank Adi Armoni for many useful discussions and the collaboration in the initial stages of the project.
I am also grateful to David Berman for the comments on a preliminary draft of the paper.
This work was supported in part by European Union's Seventh Framework Programme under grant agreements
(FP7- REGPOT-2012-2013-1) no 316165, the EU-Greece program ``Thales'' MIS 375734 and was
also co-financed by the European Union (European Social Fund, ESF) and Greek national funds
through the Operational Program ``Education and Lifelong Learning'' of the National Strategic
Reference Framework (NSRF) under ``Funding of proposals that have received a positive
evaluation in the 3rd and 4th Call of ERC Grant Schemes''.

\newpage
\begin{appendix}

\section{One M2-brane ending on $M$ M5-branes}
\label{oneM2}

As a more explicit illustration of the proposed bulk-boundary interactions, in this appendix we  
consider in more detail the interactions that are packaged in the superspace action \eqref{qftca}. 
We will focus on the relatively simpler case of a single M2-brane ending on an arbitrary 
number $M$ of M5-branes. In this case the 3d bulk ABJM theory is abelian.

\subsection*{\it 3d bulk action in $\NN=1$ superspace form}

Our starting point is the bulk action \eqref{qftae}
\bea
\label{exaa}
\SS_{bulk} &=& \SS_{CS}[V_+,V_-] + \SS_K[\phi_\pm, A, B, V_\pm] +\SS_W [\phi_\pm,A,B]
\nonumber\\
&=& \frac{k}{4\pi} \int d^3 x \, d^4\vartheta ~ \Big[ V_+ {\boldsymbol D}^\alpha \bar {\boldsymbol D}_\alpha V_+
- V_- {\boldsymbol D}^\alpha \bar {\boldsymbol D}_\alpha V_- 
\nonumber\\
&&~~~~~~~~~~~~~~~~~~~~+ \bar \varphi_+ \varphi_+ +  \bar \varphi_- \varphi_-
+\bar A_a e^{V_+} A^a e^{-V_-} + \bar B^a e^{V_-} B_a e^{-V_+} \Big]
\nonumber\\
&&+ \int d^3x\, d^2\vartheta \Big[
m (\varphi_+^2 -\varphi_-^2 ) +B_a A^a (\varphi_++\varphi_-) \Big] + {\rm c.c.}
\eea

We employ the $\NN=1$ superspace decomposition of the $\NN=2$ vector multiplets
\beq
\label{exab}
V_\pm (\theta_1,\theta_2) = \Delta_\pm (\theta_1) + \theta_2 \Gamma_{\pm}(\theta_1) 
+ \theta_2^2 \left( E_\pm(\theta_1) + D_1^2 \Delta_\pm \right) 
~,
\eeq
where $\Delta_\pm$, $E_\pm$ are $\NN=1$ real scalar superfields and $\Gamma_\pm$ are $\NN=1$ spinor
superfields. Following the conventions of \cite{Gates:1983nr} we use the notation
$D_{1\alpha}= \partial_{1\alpha} + (\gamma^\mu \theta_1)_\alpha \partial_\mu$
for the $\NN=1$ superspace derivative with respect to the real Grassmann coordinates $\theta_{1\alpha}$
($\alpha =\pm$ is a spinor index). For the $\NN=2$ chiral superfields we set
\beq
\label{exaca}
\phi_\pm(\theta_1,\theta_2) = \varphi_\pm (\theta_1) + i \theta_2 D_1 \varphi_\pm(\theta_1) 
+ \theta_2^2 D_1^2 \varphi_\pm(\theta_1)
~,
\eeq
\beq
\label{exacb}
A^a(\theta_1,\theta_2) = {\mathbb A}^a (\theta_1) + i \theta_2 D_1 {\mathbb A}^a(\theta_1) 
+ \theta_2^2 D_1^2 {\mathbb A}^a(\theta_1)
~,
\eeq
\beq
\label{exacc}
B_a(\theta_1,\theta_2) = {\mathbb B}_a (\theta_1) + i \theta_2 D_1 {\mathbb B}_a(\theta_1) 
+ \theta_2^2 D_1^2 {\mathbb B}_a(\theta_1)
~.
\eeq

Inserting these expansions in the $\NN=2$ expressions and performing the $\int d^2 \theta_2$ integrals we obtain 
in $\NN=1$ form
\bea
\label{exad}
\SS_{CS}[V_\pm]&=& \frac{k}{4\pi} \int d^3x\, d^2 \theta_1 ~ 
\left[ E_+ E_+ + \Gamma^\alpha_+ W_{+\alpha} +\frac{1}{2} D_1^\alpha \left( D_{1\alpha} E_+ \Delta_+ - 
E_+ D_{1\alpha} \Delta_+ \right) \right]
\nonumber\\
&&-  \frac{k}{4\pi} \int d^3x\, d^2 \theta_1 ~ 
\left[ E_- E_- + \Gamma^\alpha_- W_{-\alpha} +\frac{1}{2} D_1^\alpha \left( D_{1\alpha} E_- \Delta_- - 
E_- D_{1\alpha} \Delta_- \right) \right] 
~.
\eea
We remind that the gauge-invariant field strength for a spinor multiplet $\Gamma$ is 
\beq
\label{exae}
W_\alpha = \frac{1}{2} D^\beta D_\alpha \Gamma_\beta
~.
\eeq

In passing we note that it would have been considerably harder to write out the corresponding expansion
for the $\NN=2$ Chern-Simons action in general gauge in the non-abelian case. Also, note that in the so-called
Ivanov gauge one sets $\Delta_\pm=0$. This is not possible in the presence of the boundary unless we want
to start with a partially broken super-gauge symmetry.

Similarly, for the kinetic terms $S_K$ we obtain
\bea
\label{exaf}
&&\SS_K[\phi_\pm, A,B,V_\pm] = \int d^3 x \, d^2 \theta_1~ \Bigg\{ 
\frac{1}{2} D_1^\alpha \Big[
\bar \varphi_+ D_{1\alpha} \varphi_+ + D_{1\alpha} \bar \varphi_+ \varphi_+ 
+\bar \varphi_- D_{1\alpha} \varphi_- + D_{1\alpha} \bar \varphi_- \varphi_- 
\Big]
\nonumber\\
&&+\frac{1}{2} e^{\Delta_+-\Delta_-} D_1^\alpha \Big[
\bar {\mathbb A}_a D_{1\alpha} {\mathbb A}^a +D_{1\alpha} \bar {\mathbb A}_a {\mathbb A}^a \Big]
+\frac{1}{2} e^{\Delta_- -\Delta_+} D_1^\alpha \Big[
\bar {\mathbb B}^a D_{1\alpha} {\mathbb B}_a +D_{1\alpha} \bar {\mathbb B}^a {\mathbb B}_a \Big]
\nonumber\\
&&+ \Big( E_+ + E_- + D_1^2 (\Delta_+ +\Delta_- ) \Big) 
\Big( e^{\Delta_+-\Delta_-} \bar {\mathbb A}_a {\mathbb A}^a 
+e^{\Delta_- -\Delta_+} \bar {\mathbb B}^a {\mathbb B}_a \Big)
\nonumber\\
&&-2 \Big[ D_1^\alpha \bar\varphi_+ D_{1\alpha} \varphi_+
+D_1^\alpha \bar\varphi_- D_{1\alpha} \varphi_-
+e^{\Delta_+ - \Delta_-} \nabla_1^\alpha \bar {\mathbb A}_a \nabla_{1\alpha} {\mathbb A}^a
+e^{\Delta_- - \Delta_+} \nabla_1^\alpha \bar {\mathbb B}^a \nabla_{1\alpha} {\mathbb B}_a \Big]
\Bigg\}
~.
\eea
We used the $\NN=1$ super-gauge covariant derivative
\beq
\label{exag}
\nabla_{1\alpha} = D_{1\alpha} - \frac{i}{2} \left( \Gamma_{+\alpha} +\Gamma_{-\alpha} \right)
~.
\eeq

Finally, 
\bea
\label{exag}
&&\SS_W[\phi_\pm, A,B]
=\frac{1}{2}\int d^3 x \, d^2 \theta_1 ~ \Bigg\{ 
m \left( \varphi_+^2 -\varphi_-^2 \right) + 2m \theta_1 \left( D_1 \varphi_+ \varphi_+ - D_1 \varphi_- \varphi_- \right) 
\nonumber\\
&&+ m \Big( - D_1^\alpha ( \varphi_+ D_{1\alpha} \varphi_+ ) + 2 D_1 \varphi_+ D_1 \varphi_+
+D_1^\alpha ( \varphi_- D_{1\alpha} \varphi_- ) - 2 D_1 \varphi_- D_1 \varphi_- \Big)
\nonumber\\
&&+ {\mathbb B}_a {\mathbb A}^a \left( \varphi_+ + \varphi_-\right)
+2\theta_1 D_1 \Big( (\varphi_++\varphi_-) {\mathbb B}_a {\mathbb A}^a  \Big)
\nonumber\\
&&-\theta_1^2 \Big[ {\mathbb B}_a  {\mathbb A}^a D_1^2 (\varphi_+ +\varphi_- )
- D_1  {\mathbb B}_a D_1(\varphi_+ +\varphi_-)  {\mathbb A}^a
- D_1  {\mathbb A}^a D_1(\varphi_+ +\varphi_-)  {\mathbb B}_a
\nonumber\\
&&+(\varphi_+ + \varphi_-)
\Big(  {\mathbb B}_a D_1^2  {\mathbb A}^a - D_1  {\mathbb B}_a D_1  {\mathbb A}^a
+D_1^2  {\mathbb B}_a  {\mathbb A}^a \Big) \Big] \Bigg\} 
~.
\eea

We will refrain from a further evaluation of the $\theta_1$ integrals and the full expansion of these interactions in 
components. Nevertheless, it is already apparent from these expressions that there are several total-derivative 
terms that are supported on the boundary.

\subsection*{\it Boundary interactions}

We restore half of the supersymmetry by adding suitable boundary interactions to the bulk action according
to the rule \eqref{qftbb}
\beq
\label{exaba}
\SS^{(1,1)} = \int d^3x \Big\{ d^2 \theta_1 \, d^2\theta_2 \, \LL 
-d^2 \theta_1 \, \partial_2 \LL \Big |_{\theta_2=0} 
+ d^2\theta_2 \, \partial_2 \LL \Big |_{\theta_1=0}
- \partial_2 \partial_2 \LL \Big |_{\theta_1=\theta_2=0} \Big\}
~.
\eeq
In the total bulk-boundary action \eqref{qftca}
\bea
\label{exabb}
&&\SS_{proposed}[g_\pm, V_\pm, \phi_\pm, A, B] = \SS_{bulk}^{(1,1)} + \SS_{bdy}^{(gauge)} + \SS_{bdy}^{(matter)}
\nonumber\\
&&= \SS_{kin}^{(1,1)}\left[ g_+,V_+ \right] + \SS_{kin}^{(1,1)}\left[ g_-,V_- \right] 
+\SS^{(1,1)}_{CS} \left[ {\boldsymbol V}_+^{g_+}, {\boldsymbol V}_-^{g_-} \right] 
+\SS^{(1,1)}_K + \SS^{(1,1)}_W
+\SS_{bdy}^{(matter)}
\eea
the first two terms on the second line are kinetic terms on the boundary and $S_{bdy}^{matter}$ is a potential 
term on the boundary \eqref{qftbj}. The third term, $S_{CS}^{(1,1)}[{\boldsymbol V}_+^{g_+}, {\boldsymbol V}_-^{g_-}]$,
is the $\NN=2$ CS action for the non-abelian gauge group $U(M)$ with the boundary completion \eqref{exaba}.
The presence of a non-abelian boundary interaction, even for the abelian M2-brane theory, is a characteristic
difference between our proposal and previous approaches.

The remaining two terms, $\SS^{(1,1)}_K$, $\SS^{(1,1)}_W$,
on the second line of \eqref{exabb} are simple to write down. We collect the relevant expressions here.
Once again, in order to keep the expressions somewhat compact we express everything in terms 
of $\NN=1$ superfields leaving the full expansion in components, that follows straightforwardly, implicit. 
For the kinetic interactions
\bea
\label{exabc}
&&\SS^{(1,1)}_K = \int d^3 x\, d^2 \theta_1\Big[  \LL_K 
- \partial_2 \Big( \bar\varphi_+ \varphi_+ + \bar\varphi_- \varphi_- 
+ e^{\Delta_+-\Delta_-} \bar {\mathbb A}_a {\mathbb A}^a 
+ e^{\Delta_- -\Delta_+} \bar {\mathbb B}^a {\mathbb B}_a  \Big) \Big]
\nonumber\\
&&+ \int d^3 x\, \partial_2 \Big[  \LL_K
- \partial_2 \Big( \bar\varphi_+ \varphi_+ + \bar\varphi_- \varphi_- 
+ e^{\Delta_+-\Delta_-} \bar {\mathbb A}_a {\mathbb A}^a 
+ e^{\Delta_- -\Delta_+} \bar {\mathbb B}^a {\mathbb B}_a  \Big) \Big]_{\theta_1=0}
~,
\eea
where $\LL_K$ is the integrand in eq.\ \eqref{exaf}.
Finally, for the superpotential interactions
\bea
\label{exabd}
&&\SS^{(1,1)}_W = \int d^3 x\, d^2 \theta_1\Big[  \LL_W 
- \frac{1}{2} \theta_1^2 \partial_2 \Big( m  (\varphi_+^2 -\varphi_-^2) 
+ {\mathbb B}_a {\mathbb A}^a (\varphi_+ +\varphi_-) \Big) \Big]
\nonumber\\
&&+\frac{1}{2} \int d^3 x \, \partial_2 \Big[
m (\varphi_+^2 -\varphi_-^2) 
+ m \Big( - D_1^\alpha (\varphi_+ D_{1\alpha} \varphi_+) + 2 D_1 \varphi_+ D_1\varphi_+
\nonumber\\
&&+ D_1^\alpha (\varphi_- D_{1\alpha} \varphi_-) - 2 D_1 \varphi_- D_1\varphi_- \Big)
+{\mathbb B}_a  {\mathbb A}^a (\varphi_+ + \varphi_-) \Big]_{\theta_1=0} +{\rm c.c.}
~,
\eea
where $\LL_W$ is the integrand in eq.\ \eqref{exag}.

\end{appendix}

\newpage
\providecommand{\href}[2]{#2}\begingroup\raggedright

\end{document}